\documentstyle[11pt]{article}

\newcommand{\bi}{\mbox{\boldmath $i$}}
\newcommand{\bj}{\mbox{\boldmath $j$}}
\newcommand{\bl}{\mbox{\boldmath $l$}}
\newcommand{\bq}{\mbox{\boldmath $q$}}
\newcommand{\bk}{\mbox{\boldmath $k$}}
\begin{document}
\title{Glassy Transition in the three-dimensional\\
Random Field Ising Model\\}
\author{M. M\'ezard $^*$, R. Monasson $^{**}$\\
\\
$^*$ Laboratoire de Physique Th\'eorique de l'Ecole Normale Sup\'erieure
$^{(1)}$,
\\
24 rue Lhomond,  75231 Paris Cedex 05, France \\
\\
$^{**}$ INFN - Sezione di Roma I \\
c/o Dipartimento di Fisica, Universit\`a di Roma ``La Sapienza'',\\
Piazzale A. Moro 2, 00185 Roma, Italy \\}
\footnotetext[1]{Unit\'e propre de recherche du CNRS, associ\'ee \`a
l'Universit\'e  de Paris-Sud et \`a l'ENS. }

\maketitle

PACS numbers~: 05.20 - Statistical Mechanics
\hfill \break \hspace*{3.45 truecm}
64.60 - Order-Disorder and Statistical Mechanics of Model Systems

\vskip 1cm

\begin{abstract}
The high temperature phase of the three dimensional random field Ising model
is studied using replica symmetry breaking framework. It is found
that, above the ferromagnetic transition temperature $T_f$,
there appears a glassy phase at intermediate temperatures $T_f<T<T_b$
while the usual paramagnetic phase exists for $T>T_b$ only.
Correlation length at $T_b$ is computed and found to be
compatible with previous numerical results.
\end{abstract}

\vskip 1cm

Although a great deal of work has been devoted to the understanding of the
random field Ising model (RFIM) \cite{rev}, some aspects still need to be
cleared up. It is now well-known that, in dimension $D=3$, long
range order is present at sufficiently low temperature and weak random
fields with non trivial critical exponents
\cite{lro}, the upper critical dimension
of the RFIM being $D=6$. Nevertheless, perturbation theory leads to
dimensional reduction (critical exponents are incorrectly predicted to be
equal to those of the corresponding pure model in dimension $D-2$) \cite{dr}
and therefore does not succeed in describing the critical behaviour of the
RFIM. The reason of this failure presumably stems from the very complicated
energy landscape due to the quenched disorder, and more precisely, from the
existence of a huge number of local minima of the free energy in the space
of local magnetisations that usual perturbative expansions do not
take into account \cite{lot}. Numerical simulations and resolutions of the
mean-field equations corroborate this picture \cite{oldsim,newsim}. Above
the ferromagnetic transition temperature $T_f$, there seems to appear an
intermediate ``glassy''
regime for $T_f < T< T_b$ where many solutions of the local
magnetisations mean-field equations coexist, while only one of them subsists
in the paramagnetic phase $T>T_b$. From the theoretical point of view, it
was suggested that the techniques of replica symmetry breaking (RSB), which
proved to be successful in the mean-field theory of spin glasses \cite{mpv}
where such complicated free energy landscapes arise, could also be applied
to the RFIM \cite{oldrsb}. Experiments made on diluted
anti-ferromagnets also found an irreversibility line above the critical
temperature where the anti-ferromagnetic order appears \cite{bel91}.
Recently, M\'ezard and Young, referred to
in the following as M$-$Y, proposed a
variational approach  of the RFIM \cite{my} based on a self-consistent
expansion in $1/N$ (where $N$ is the number of spin components) due to
Bray \cite{bray}. They found that replica symmetry, which gives back
dimensional reduction, must be broken at the ferromagnetic transition
$T=T_f$ and
that RSB solution leads to sensible results for the critical exponents
in agreement with already known results \cite{rev,my}.

In this paper, using M$-$Y's framework,
we concentrate upon the non ferromagnetic regime (i.e. $T> T_f$). We find
that there exist indeed two different phases~: a paramagnetic phase at
high temperatures $T>T_b$ and a glassy phase at intermediate temperatures
$T_f<T<T_b$. The value of the correlation length at $T=T_b$ where
the RSB transition occurs is computed and
compared to predictions obtained from numerical resolution of mean-field
equations \cite{newsim}.

The model we consider is a $N$-component version of the RFIM on a
three-dimensional lattice including $L^3$ spins ${\bf \Phi} _{\bi}
= (\Phi ^1 _{\bi}, ..., \Phi ^N _{\bi})$, where $\bi =(
i_1, i_2, i_3)$ and $0\le i_1, i_2, i_3 \le L-1$,
\begin{equation}
\label{dh}
{\cal H}({\bf \Phi},{\bf h}) = {1\over 2} \sum _{\langle \bi ,\bj \rangle}
\left( {\bf \Phi _{\bi}} - {\bf \Phi _{\bj}} \right) ^2 + {r \over 2}
\sum _{\bi} \left( {\bf \Phi} _{\bi} \right) ^2
+ {1\over 4N} \sum _{\bi} \left( \left( {\bf \Phi} _{\bi} \right) ^2 \right) ^2
- \sum _{\bi} {\bf h} _{\bi} . {\bf \Phi _{\bi}}
\end{equation}
where ${\bf h} _{\bi}$ is a quenched random field, the distribution of
which is  Gaussian, uncorrelated at differents sites, with mean
$\overline{h ^{\mu} _{\bi}}=0$ and variance $\overline{h ^{\mu} _{\bi}.
h ^{\lambda} _{\bj}}= \Delta \ \delta ^{\mu \lambda}\ \delta _{\bi \bj}$.
Following the standard procedure \cite{mpv}, we introduce
$n$ replicas of the spins ${\bf \Phi} ^a$, $a=1...n$, and average over the
quenched disorder $\bf h$ to obtain the effective Hamiltonian
\begin{equation}
\label{ef}
{\cal H}(\{{\bf \Phi}^a\})={1 \over 2} \sum _{\langle \bi
,\bj \rangle , a}
\left( {\bf \Phi}^a _{\bi} - {\bf \Phi}^a _{\bj} \right) ^2 + {r \over 2}
\sum _{\bi ,a} \left( {\bf \Phi} ^a_{\bi} \right)^2
- {\Delta \over 2} \sum _{\bi , (a,b)}{\bf \Phi}^a_{\bi}.{\bf \Phi}^b _{\bi}
+ {1\over 4N} \sum _{\bi ,a} \left( \left(
{\bf \Phi}^a _{\bi} \right) ^2 \right) ^2
\end{equation}
The replica correlation functions $\langle \Phi ^{\mu ,a} _{\bk} \Phi
^{\lambda ,b}
_{-\bk} \rangle = \delta ^{\mu \lambda} \ G^{ab}(\bk )$ with the effective
Hamiltonian (\ref{ef}) are related to the disconnected and connected
correlation functions, $\overline{\langle \Phi _{\bk} ^{\mu} \rangle _h .
\langle \Phi _{-\bk} ^{\lambda} \rangle _h} = \delta ^{\mu \lambda}
\ G_{dis}(\bk)$ and $\overline{\langle \Phi _{\bk} ^{\mu}.\Phi _{-\bk}
^{\lambda} \rangle _h -\langle \Phi _{\bk} ^{\mu} \rangle _h .
\langle \Phi _{-\bk} ^{\lambda} \rangle _h} =\delta ^{\mu \lambda}
\ G_{con}(\bk)$, where $\langle . \rangle _h$ denotes the average over the
Gibbs measure induced by (\ref{dh}) \cite{my}. For
the replica symmetric assumption $G^{ab}(\bk )= \tilde G (\bk )\ \delta ^{ab}
+G(\bk )$, the correspondance is simply $G_{dis}(\bk )=G(\bk )$ and
$G_{con}(\bk )=\tilde G(\bk )$.

Using Bray's self-consistent screening approximation \cite{bray} which is
exact to order $1/N$, one finds that the propagators $G^{ab}(\bk )$ are
given by the saddle-point of the free energy
\begin{eqnarray}
\label{energlibre}
{\cal F}\left( \{ G^{ab} (\bk) \} \right) &=& \sum _{\bk ,a} \left(
6 - 2 \sum _{d=1}^3 \cos \left({2 \pi k _d \over L} \right) + r +
{1 \over 2 L^3} \sum _{\bq} G^{aa}(\bq ) - \Delta \right)
G^{aa}(\bk ) \nonumber \\ &-& 2 \Delta \sum _{\bk , a<b} G^{ab}(\bk )
- \sum _{\bk ,a} \bigg( \log G(\bk ) \bigg) ^{aa} + {1\over N}
\sum _{\bk ,a} \bigg( \log \left[ 1+\Pi (\bk ) \right] \bigg) ^{aa}
\end{eqnarray}
where
\begin{equation}
\label{piab}
\Pi ^{ab}(\bk )= {1\over L^3} \sum _{\bq} G^{ab}(\bk - \bq )\ G^{ab}(\bq )
\end{equation}
Within the replica symmetric hypothesis, the correlation functions
are therefore solutions of the following set of $2 L^3$ implicit equations
\begin{eqnarray}
\label{saddle1}
{G(\bk ) \over [\tilde G( \bk )]^2} &=& \Delta + {2\over N L^3} \sum _{\bq}
{G(\bk - \bq)\ \Pi (\bq) \over [ 1+ \tilde \Pi (\bq )]^2 } \\
\label{saddle2}
{1\over \tilde G(\bk )} &=& 6 - 2 \sum _{d=1}^3 \cos \left({2 \pi k _d
\over L} \right) + r + {1 \over L^3} \sum _{\bq } \bigg( \tilde G(\bq ) +
G(\bq ) \bigg) \nonumber \\
&+& {2\over N L^3} \sum _{\bq} \left[ {\tilde G(\bk -\bq )+G(\bk -\bq ) \over
1+\tilde \Pi (\bq)} - {\tilde G(\bk - \bq)\ \Pi (\bq) \over
[ 1+ \tilde \Pi (\bq )]^2 } \right]
\end{eqnarray}
where
\begin{eqnarray}
\label{pirs}
\tilde \Pi (\bk )&=& {1\over L^3} \sum _{\bq} \bigg[ \tilde G (\bk - \bq )\
\tilde G(\bq ) + G (\bk - \bq )\ \tilde G(\bq )+\tilde G (\bk - \bq )
\ G(\bq ) \bigg] \\
\Pi (\bk )&=& {1\over L^3} \sum _{\bq} G (\bk - \bq )\ G(\bq )
\end{eqnarray}

In order to determine where the transition to RSB occurs, we have studied
the local stability of the free energy ${\cal F}(G^{ab})$
around the symmetric saddle-point $(\tilde G ,G)$. Repeating de~Almeida$-$
Thouless's calculations \cite{dat} and taking into account the
$\bk$-dependence of the order parameters $\tilde G$ and $G$, we have found
that the replica symmetric solution is locally stable if and only if the
lowest eigenvalue $\Lambda$ of the matrix
\begin{equation}
\label{matrix}
{\cal M}(\bk ,\bl ) = {\delta _{\bk ,\bl } \over [\tilde G(\bk )]^2} -
{2 \over N L^3} {\Pi (\bk -\bl ) \over [1+\tilde \Pi (\bk -\bl )]^2 } -
{4 \over N L^6} \sum _{\bq} {G(\bk -\bq )\ G(\bl -\bq ) \over [1+\tilde
\Pi (\bq )]^2 }
\end{equation}
is strictly positive.

When $N\to \infty$, Bray's partial resummation reduces to the Hartree-Fock
approximation. $\tilde G(\bk )$ is thus equal to the bare propagator with
a renormalised squared mass $\tilde m^2$ solution of the gap
equation (\ref{saddle1},
\ref{saddle2}). From the expression of the AT matrix (\ref{matrix}), one
obtains $\Lambda =\tilde m^4 > 0$. The replica symmetric solution is therefore
always stable. As soon as $N$ becomes finite, the corrections appearing in
(\ref{matrix})  may lead to instabilities.
In our three-dimensional system, however, the self-consistent screening
approximation induces no ferromagnetic transition for large $N$
($\tilde m$ never vanishes for finite
bare temperatures $r$) \cite{my,bray}. Hereafter, we choose $N=1$
(Ising case) which allows for the existence of long range order at finite
$r<0$.

For every size $L$ of the lattice, we fix a value of $r$ and solve
for the propagators $\tilde G, G$ by an iteration of
equations (\ref{saddle1}, \ref{saddle2}). Using rotational and translational
symmetries,
only $\tilde G(\bk )$ and $G(\bk )$ with $0\le k_1
\le k_2 \le k_3 \le Int({L\over 2})$ are to be found. Once a
fixed point is reached, we estimate the mass
$\tilde m$ and the correlation length $\tilde \xi = {1\over \tilde m}$
from the low-momentum behaviour of $\tilde G$
\begin{equation}
\label{fitm}
\tilde G (0,0,0) \simeq  {\tilde a \over \tilde m ^2} \qquad , \qquad
\tilde G (0,0,1) \simeq  {\tilde a \over 2-2\cos({2 \pi \over L})+ \tilde m ^2}
\end{equation}
\footnotetext[2]{For e.g. $N=4$, long range order is absent and both mass
$\tilde m$ and eigenvalue $\Lambda$ are always non zero. However,
from numerical resolution of the saddle-point equations (\ref{saddle1},
\ref{saddle2}), we found that using definition (\ref{fitm}) of $\tilde
m$, the relation $\Lambda \simeq \tilde m ^4$ is roughly correct. }
Expression (\ref{fitm}) is exact for $N = \infty ^{\ (2)}$. Moreover,
from perturbation
theory which is thought to be correct above the $AT$ transition, we expect
$\tilde G(\bk )$ to have a single pole. A similar calculation gives
$m$ and $\xi ={1\over m}$, assuming that $G(\bk )$ has a double pole $^{(3)}$.
The lowest eigenvalue $\Lambda$ is then computed by diagonalising the
AT matrix. This highly time consuming task may be simplified by observing
that ${\cal M}$ is invariant under the three symmetry operators ${\cal
S}_d : k_d \to
L - k_d$. In the base of their eigenvectors, ${\cal M}$ reduces to eight
diagonal blocs of size roughly equal to $\left( {L\over 2}\right) ^3 \times
\left( {L\over 2}\right) ^3$. On can check that the eigenvector corresponding
to $\Lambda$ belongs to the ``physical'' subspace, i.e. the one spanned by
eigenvectors of ${\cal S}_d$ of eigenvalues $+1$.
The process is repeated until the value $r_L$ of the bare temperature $r$
where $\Lambda$ vanishes and the corresponding
correlation lengths $\tilde \xi _L$ and $\xi _L$ are bracketed with
a sufficient precision. The final uncertainties on $\tilde \xi _L$ and
$\xi _L$ are lower than $\pm 5.10^{-4}$ for lattice sizes running from $L=2$
up to $L=20$.

The numerical values of the correlations lengths at the AT transition are
displayed Fig.~1. Although it seems difficult to extrapolate to $L
\to \infty$, reliable information on the thermodynamical limit
may be obtained since the correlations lengths
are relatively small as compared to the lattice size ($\tilde \xi _L <
\xi _L < L/3$ for $L=20$). From finite size effect theory \cite{size},
we expect indeed that, if the mass $\tilde m _L$ converges to a finite
value $\tilde m _{\infty}>0$ at the thermodynamical limit, then its
asymptotic behaviour obeys
\begin{equation}
\tilde m_L - \tilde m_{\infty}\ \simeq \ \tilde C \ .\ e^{-L/\tilde \xi _L} +
O \left(e^{- 2L/\tilde \xi _L} \right)
\label{sca}
\end{equation}
where $\tilde C$ is a constant (the same identity holds for $m$ and
$\xi _L$ with a different constant $C$). Fig.~2 shows the dependence of
$\tilde m _L$ and $m _L$ upon $e^{-L/\tilde \xi _L}$ and $e^{-L/\xi _L}$
respectively. The linear law (\ref{sca}) is very well verified with
proportionality factors of order one ($\tilde C \simeq 1.83$, $C \simeq
1.37$) $^{(4)}$. Linear extrapolations to $L \to \infty$ provide the
values of the correlation lengths at the thermodynamical limit
\begin{equation}
\tilde \xi _{\infty}\ \simeq \ \xi _{\infty} \ \simeq \ 7.7 \pm 0.2
\label{limit}
\end{equation}
The equality between the correlation lengths defined from the disconnected
and the connected correlation functions is a self-consistent check of our
analytical and numerical results. It is indeed predicted by perturbation
theory \cite{dr,newsim} and therefore holds for high temperatures down to
the RSB transition.

\footnotetext[3]{This stems from perturbation theory which is thought
to be valid above $T_b$ \cite{dr} and can be checked up when $N \to
\infty$ on (\ref{saddle1}, \ref{saddle2}). For the Ising system, the one
dimensional correlations functions in the real space, $\tilde
 g(x)$ and $g(x)$ (which are respectively the Fourier transforms of
$\tilde G(k,0,0)$ and $G(k,0,0)$) can be fitted with a very good agreement
by $\tilde g(x)\simeq \tilde a \cosh (x/\tilde \xi) +\tilde b$ and
$g(x)\simeq a (1+ x/\xi)\cosh (x/ \xi) + b$ \cite{newsim}. The discrepancy
between these values of the correlation lengths and the ones defined in the
text seems to vanish for increasing lattice sizes.}

\footnotetext[4]{Although scaling relation (\ref{sca}) should also be valid for
the correlation length itself, i.e. $\tilde \xi _{\infty} - \tilde
\xi _L \ \simeq \ \tilde D \ .\ e^{-L/\tilde \xi _L}$, one sees from
$\tilde D = - \tilde C .\ \xi _{\infty} ^2 \simeq - 100$ that it
cannot be directly observed with lattice sizes lower than $L=20$.}

In this letter, we have argued that the non-ferromagnetic phase of the
three-dimensional random field Ising model is composed of a paramagnetic
phase at high temperatures and a spin-glass phase at lower temperatures. The
onset of this glassy phase therefore occurs at a finite correlation length
for both correlation and susceptibility functions
which was found to be in the range $7.5 < \xi < 8$. Although such a
result might be an artefact due to the $1/N$ approach used here, it is
in qualitative agreement with previous numerical studies which found that
the mean field equations of the RFIM begin to have more than one
solution, and thus that the perturbative approach ceases to be
correct, for $\xi > 4.5$. Such a behaviour is to be expected in the low
temperature phase too. Since at very low $T$, only
the two states where all spins are aligned along the same direction remain,
replica symmetry has to be restored at a temperature $T_s$ with $0<T_s<
T_f$. It would be interesting to extend the calculation we have presented
here to verify explicitely this conjecture.

\vskip .5cm
We are particularly grateful to G. Parisi for numerous and fruitful
discussions and suggestions.
We also wish to thank M. Aizenman, C. De Dominicis, T. Garel, D.J. Lancaster
and H. Orland
for interesting discussions. This study was partly funded by the EEC and the
French Ministry of Research.

\section*{Figure Captions}

Figure 1~: the correlation lengths $\tilde \xi _L$ (empty dots) and $\xi _L$
(full dots) at the
onset on the spin glass phase for different lattice sizes $L=2$ up to
$L=20$ (the total number of spins is $L^3$).
\vskip .1cm \hfill \break
Figure 2~: the masses $\tilde m _L= {1\over \tilde \xi _L}$ (empty dots) and
$m_L={1\over \xi _L}$ (full dots) plotted vs. the ``finite size'' factors
$\exp (-L/\tilde \xi _L)$ and $\exp (-L/ \xi _L)$ respectively. The dashed
lines
are the best linear fits from the last twelve points (sizes $L=9$ to $L=20$).

\begin{figure}[p]

\setlength{\unitlength}{0.240900pt}
\ifx\plotpoint\undefined\newsavebox{\plotpoint}\fi
\begin{picture}(1500,900)(0,0)
\font\gnuplot=cmr10 at 10pt
\gnuplot
\sbox{\plotpoint}{\rule[-0.200pt]{0.400pt}{0.400pt}}%
\put(176.0,68.0){\rule[-0.200pt]{0.400pt}{194.888pt}}
\put(176.0,68.0){\rule[-0.200pt]{4.818pt}{0.400pt}}
\put(154,68){\makebox(0,0)[r]{1.5}}
\put(1416.0,68.0){\rule[-0.200pt]{4.818pt}{0.400pt}}
\put(176.0,158.0){\rule[-0.200pt]{4.818pt}{0.400pt}}
\put(154,158){\makebox(0,0)[r]{2}}
\put(1416.0,158.0){\rule[-0.200pt]{4.818pt}{0.400pt}}
\put(176.0,248.0){\rule[-0.200pt]{4.818pt}{0.400pt}}
\put(154,248){\makebox(0,0)[r]{2.5}}
\put(1416.0,248.0){\rule[-0.200pt]{4.818pt}{0.400pt}}
\put(176.0,338.0){\rule[-0.200pt]{4.818pt}{0.400pt}}
\put(154,338){\makebox(0,0)[r]{3}}
\put(1416.0,338.0){\rule[-0.200pt]{4.818pt}{0.400pt}}
\put(176.0,428.0){\rule[-0.200pt]{4.818pt}{0.400pt}}
\put(154,428){\makebox(0,0)[r]{3.5}}
\put(1416.0,428.0){\rule[-0.200pt]{4.818pt}{0.400pt}}
\put(176.0,517.0){\rule[-0.200pt]{4.818pt}{0.400pt}}
\put(154,517){\makebox(0,0)[r]{4}}
\put(1416.0,517.0){\rule[-0.200pt]{4.818pt}{0.400pt}}
\put(176.0,607.0){\rule[-0.200pt]{4.818pt}{0.400pt}}
\put(154,607){\makebox(0,0)[r]{4.5}}
\put(1416.0,607.0){\rule[-0.200pt]{4.818pt}{0.400pt}}
\put(176.0,697.0){\rule[-0.200pt]{4.818pt}{0.400pt}}
\put(154,697){\makebox(0,0)[r]{5}}
\put(1416.0,697.0){\rule[-0.200pt]{4.818pt}{0.400pt}}
\put(176.0,787.0){\rule[-0.200pt]{4.818pt}{0.400pt}}
\put(154,787){\makebox(0,0)[r]{5.5}}
\put(1416.0,787.0){\rule[-0.200pt]{4.818pt}{0.400pt}}
\put(176.0,877.0){\rule[-0.200pt]{4.818pt}{0.400pt}}
\put(154,877){\makebox(0,0)[r]{6}}
\put(1416.0,877.0){\rule[-0.200pt]{4.818pt}{0.400pt}}
\put(176.0,68.0){\rule[-0.200pt]{0.400pt}{4.818pt}}
\put(176,23){\makebox(0,0){0}}
\put(176.0,857.0){\rule[-0.200pt]{0.400pt}{4.818pt}}
\put(462.0,68.0){\rule[-0.200pt]{0.400pt}{4.818pt}}
\put(462,23){\makebox(0,0){5}}
\put(462.0,857.0){\rule[-0.200pt]{0.400pt}{4.818pt}}
\put(749.0,68.0){\rule[-0.200pt]{0.400pt}{4.818pt}}
\put(749,23){\makebox(0,0){10}}
\put(749.0,857.0){\rule[-0.200pt]{0.400pt}{4.818pt}}
\put(1035.0,68.0){\rule[-0.200pt]{0.400pt}{4.818pt}}
\put(1035,23){\makebox(0,0){15}}
\put(1035.0,857.0){\rule[-0.200pt]{0.400pt}{4.818pt}}
\put(1321.0,68.0){\rule[-0.200pt]{0.400pt}{4.818pt}}
\put(1321,23){\makebox(0,0){20}}
\put(1321.0,857.0){\rule[-0.200pt]{0.400pt}{4.818pt}}
\put(176.0,68.0){\rule[-0.200pt]{303.534pt}{0.400pt}}
\put(1436.0,68.0){\rule[-0.200pt]{0.400pt}{194.888pt}}
\put(176.0,877.0){\rule[-0.200pt]{303.534pt}{0.400pt}}
\put(176.0,68.0){\rule[-0.200pt]{0.400pt}{194.888pt}}
\put(291,109){\circle*{18}}
\put(348,160){\circle*{18}}
\put(405,214){\circle*{18}}
\put(462,264){\circle*{18}}
\put(520,313){\circle*{18}}
\put(577,358){\circle*{18}}
\put(634,402){\circle*{18}}
\put(691,443){\circle*{18}}
\put(749,482){\circle*{18}}
\put(806,520){\circle*{18}}
\put(863,556){\circle*{18}}
\put(921,590){\circle*{18}}
\put(978,623){\circle*{18}}
\put(1035,654){\circle*{18}}
\put(1092,685){\circle*{18}}
\put(1150,713){\circle*{18}}
\put(1207,741){\circle*{18}}
\put(1264,768){\circle*{18}}
\put(1321,793){\circle*{18}}
\put(291,135){\circle{18}}
\put(348,192){\circle{18}}
\put(405,250){\circle{18}}
\put(462,304){\circle{18}}
\put(520,354){\circle{18}}
\put(577,402){\circle{18}}
\put(634,446){\circle{18}}
\put(691,489){\circle{18}}
\put(749,529){\circle{18}}
\put(806,567){\circle{18}}
\put(863,603){\circle{18}}
\put(921,638){\circle{18}}
\put(978,671){\circle{18}}
\put(1035,702){\circle{18}}
\put(1092,732){\circle{18}}
\put(1150,761){\circle{18}}
\put(1207,788){\circle{18}}
\put(1264,814){\circle{18}}
\put(1321,839){\circle{18}}
\end{picture}

\caption{correlation lengths $\tilde \xi _L$ (empty dots) and $\xi _L$
(full dots) at the RSB transition for differents lattice sizes $L$.}
\vspace{2cm}

\setlength{\unitlength}{0.240900pt}
\ifx\plotpoint\undefined\newsavebox{\plotpoint}\fi
\sbox{\plotpoint}{\rule[-0.200pt]{0.400pt}{0.400pt}}%
\begin{picture}(1500,900)(0,0)
\font\gnuplot=cmr10 at 10pt
\gnuplot
\sbox{\plotpoint}{\rule[-0.200pt]{0.400pt}{0.400pt}}%
\put(176.0,68.0){\rule[-0.200pt]{303.534pt}{0.400pt}}
\put(176.0,68.0){\rule[-0.200pt]{0.400pt}{194.888pt}}
\put(176.0,68.0){\rule[-0.200pt]{4.818pt}{0.400pt}}
\put(154,68){\makebox(0,0)[r]{0}}
\put(1416.0,68.0){\rule[-0.200pt]{4.818pt}{0.400pt}}
\put(176.0,169.0){\rule[-0.200pt]{4.818pt}{0.400pt}}
\put(154,169){\makebox(0,0)[r]{0.1}}
\put(1416.0,169.0){\rule[-0.200pt]{4.818pt}{0.400pt}}
\put(176.0,270.0){\rule[-0.200pt]{4.818pt}{0.400pt}}
\put(154,270){\makebox(0,0)[r]{0.2}}
\put(1416.0,270.0){\rule[-0.200pt]{4.818pt}{0.400pt}}
\put(176.0,371.0){\rule[-0.200pt]{4.818pt}{0.400pt}}
\put(154,371){\makebox(0,0)[r]{0.3}}
\put(1416.0,371.0){\rule[-0.200pt]{4.818pt}{0.400pt}}
\put(176.0,473.0){\rule[-0.200pt]{4.818pt}{0.400pt}}
\put(154,473){\makebox(0,0)[r]{0.4}}
\put(1416.0,473.0){\rule[-0.200pt]{4.818pt}{0.400pt}}
\put(176.0,574.0){\rule[-0.200pt]{4.818pt}{0.400pt}}
\put(154,574){\makebox(0,0)[r]{0.5}}
\put(1416.0,574.0){\rule[-0.200pt]{4.818pt}{0.400pt}}
\put(176.0,675.0){\rule[-0.200pt]{4.818pt}{0.400pt}}
\put(154,675){\makebox(0,0)[r]{0.6}}
\put(1416.0,675.0){\rule[-0.200pt]{4.818pt}{0.400pt}}
\put(176.0,776.0){\rule[-0.200pt]{4.818pt}{0.400pt}}
\put(154,776){\makebox(0,0)[r]{0.7}}
\put(1416.0,776.0){\rule[-0.200pt]{4.818pt}{0.400pt}}
\put(176.0,877.0){\rule[-0.200pt]{4.818pt}{0.400pt}}
\put(154,877){\makebox(0,0)[r]{0.8}}
\put(1416.0,877.0){\rule[-0.200pt]{4.818pt}{0.400pt}}
\put(176.0,68.0){\rule[-0.200pt]{0.400pt}{4.818pt}}
\put(176,23){\makebox(0,0){0}}
\put(176.0,857.0){\rule[-0.200pt]{0.400pt}{4.818pt}}
\put(356.0,68.0){\rule[-0.200pt]{0.400pt}{4.818pt}}
\put(356,23){\makebox(0,0){0.05}}
\put(356.0,857.0){\rule[-0.200pt]{0.400pt}{4.818pt}}
\put(536.0,68.0){\rule[-0.200pt]{0.400pt}{4.818pt}}
\put(536,23){\makebox(0,0){0.1}}
\put(536.0,857.0){\rule[-0.200pt]{0.400pt}{4.818pt}}
\put(716.0,68.0){\rule[-0.200pt]{0.400pt}{4.818pt}}
\put(716,23){\makebox(0,0){0.15}}
\put(716.0,857.0){\rule[-0.200pt]{0.400pt}{4.818pt}}
\put(896.0,68.0){\rule[-0.200pt]{0.400pt}{4.818pt}}
\put(896,23){\makebox(0,0){0.2}}
\put(896.0,857.0){\rule[-0.200pt]{0.400pt}{4.818pt}}
\put(1076.0,68.0){\rule[-0.200pt]{0.400pt}{4.818pt}}
\put(1076,23){\makebox(0,0){0.25}}
\put(1076.0,857.0){\rule[-0.200pt]{0.400pt}{4.818pt}}
\put(1256.0,68.0){\rule[-0.200pt]{0.400pt}{4.818pt}}
\put(1256,23){\makebox(0,0){0.3}}
\put(1256.0,857.0){\rule[-0.200pt]{0.400pt}{4.818pt}}
\put(1436.0,68.0){\rule[-0.200pt]{0.400pt}{4.818pt}}
\put(1436,23){\makebox(0,0){0.35}}
\put(1436.0,857.0){\rule[-0.200pt]{0.400pt}{4.818pt}}
\put(176.0,68.0){\rule[-0.200pt]{303.534pt}{0.400pt}}
\put(1436.0,68.0){\rule[-0.200pt]{0.400pt}{194.888pt}}
\put(176.0,877.0){\rule[-0.200pt]{303.534pt}{0.400pt}}
\put(176.0,68.0){\rule[-0.200pt]{0.400pt}{194.888pt}}
\sbox{\plotpoint}{\rule[-0.500pt]{1.000pt}{1.000pt}}%
\put(1309,653){\circle*{18}}
\put(988,570){\circle*{18}}
\put(813,506){\circle*{18}}
\put(699,458){\circle*{18}}
\put(618,421){\circle*{18}}
\put(557,393){\circle*{18}}
\put(508,369){\circle*{18}}
\put(469,350){\circle*{18}}
\put(436,334){\circle*{18}}
\put(408,320){\circle*{18}}
\put(385,308){\circle*{18}}
\put(364,298){\circle*{18}}
\put(346,288){\circle*{18}}
\put(330,280){\circle*{18}}
\put(316,273){\circle*{18}}
\put(304,267){\circle*{18}}
\put(292,261){\circle*{18}}
\put(282,256){\circle*{18}}
\put(273,251){\circle*{18}}
\put(1414,608){\circle{18}}
\put(1091,530){\circle{18}}
\put(907,471){\circle{18}}
\put(784,428){\circle{18}}
\put(693,395){\circle{18}}
\put(623,369){\circle{18}}
\put(567,349){\circle{18}}
\put(521,331){\circle{18}}
\put(483,317){\circle{18}}
\put(451,305){\circle{18}}
\put(423,294){\circle{18}}
\put(398,285){\circle{18}}
\put(377,276){\circle{18}}
\put(358,269){\circle{18}}
\put(341,263){\circle{18}}
\put(326,257){\circle{18}}
\put(313,252){\circle{18}}
\put(301,247){\circle{18}}
\put(290,243){\circle{18}}
\put(280,239){\circle{18}}
\put(176,199){\usebox{\plotpoint}}
\multiput(176,199)(19.386,7.416){65}{\usebox{\plotpoint}}
\put(1436,681){\usebox{\plotpoint}}
\put(176,199){\usebox{\plotpoint}}
\multiput(176,199)(18.440,9.527){69}{\usebox{\plotpoint}}
\put(1436,850){\usebox{\plotpoint}}
\end{picture}

\caption{masses $\tilde m _L = {1\over \tilde \xi _L}$ (empty dots) and
$m_L={1\over \xi _L}$ (full dots) vs.
$e^ {-L/\tilde \xi _L}$ and $e^{ -L/ \xi _L}$ respectively. The dashed lines
are the best linear fits from the last twelve points (sizes $L=9$ to $L=20$).}
\end{figure}

\end{document}